\documentclass[aps,prb,two column,superscriptaddress,amsmath,amssymb]{revtex4-2}
\usepackage{graphicx}
\usepackage{epsfig}
\usepackage{epstopdf}
\usepackage{amsmath}
\usepackage{mathrsfs}
\usepackage{xcolor}
\usepackage[normalem]{ulem}

\usepackage{hyperref}
\hypersetup{
    colorlinks=true,
    linkcolor=blue,
    citecolor=blue,
    filecolor=magenta,      
    urlcolor=blue,
    }

\begin{document}
\setcitestyle{super}

\title{Why Fe Doping Kills Photoluminescence in CsPbCl$_3$ but Not in CsPbBr$_3$: Role of Midgap Fe 3$d$ States and Electron-Phonon Coupling}

\author{Arpan Das}
\email{arpandas9236@gmail.com}
\affiliation{Theoretical Sciences Unit, Jawaharlal Nehru Centre for Advanced Scientific Research, Jakkur, Bangalore 560064, India}
\affiliation{Department of Science and Humanities, Audisankara (Deemed to be University), Gudur 524101, India}

\author{Saptarshi Chakraborty}
\altaffiliation{Present Address: Laboratory for Energy Materials, École Polytechnique Fédérale de Lausanne, CH-1015, Lausanne, Switzerland}
\affiliation{New Chemistry Unit, Jawaharlal Nehru Centre for Advanced Scientific Research, Jakkur, Bangalore 560064, India}

\author{Ranjani Viswanatha}
\affiliation{New Chemistry Unit, Jawaharlal Nehru Centre for Advanced Scientific Research, Jakkur, Bangalore 560064, India}
\affiliation{International Centre for Materials Science, Jawaharlal Nehru Centre for Advanced Scientific Research, Jakkur, Bangalore 560064, India}

\author{Shobhana Narasimhan}
\email{shobhana@jncasr.ac.in}
\affiliation{Theoretical Sciences Unit, Jawaharlal Nehru Centre for Advanced Scientific Research, Jakkur, Bangalore 560064, India}
\affiliation{International Centre for
Theoretical Sciences, Hessaraghatta Hobli, Bangalore
560089, India}


\begin{abstract}

Understanding the impact of transition-metal doping on the optoelectronic properties of halide perovskite nanocrystals is essential for their rational design in photonic applications. We establish the microscopic origin of photoluminescence (PL) quenching in Fe-doped CsPbCl$_3$ using spin-polarized density functional theory calculations. The emergence of Fe 3$d$ midgap states creates efficient electron-trapping centres that drive nonradiative recombination, accounting for the reduced PL intensity. Extending this analysis to Fe-doped CsPbX$_3$ (X = Cl, Br), we show experimentally that although PL intensity is suppressed in both systems relative to their pristine counterparts, their high-doping behaviour diverges: CsPbCl$_3$ becomes completely non-emissive, whereas CsPbBr$_3$ retains a finite, saturated PL intensity. Despite this contrast, electronic structure calculations reveal nearly identical midgap states in both materials, indicating that electronic effects alone cannot explain the distinct PL responses. Phonon calculations likewise fail to capture this difference. In contrast, electron-phonon coupling calculations based on the deformation potential approach reveal significantly stronger coupling in Fe-doped CsPbCl$_3$, enabling efficient dissipation of electronic excitation energy into lattice vibrations and leading to complete PL quenching. These results identify electron-phonon coupling as the key factor governing halide-dependent PL quenching and provide a unified microscopic framework for dopant-induced nonradiative processes in halide perovskites.

\end{abstract}

\maketitle

\section{INTRODUCTION}

Halide perovskites have attracted significant attention due to their remarkable optoelectronic properties and wide-ranging applications in photovoltaics, light-emitting devices, photodetectors, and lasers. These materials, with the general formula ABX$_3$, exhibit tunable band gaps, high photoluminescence quantum yields (PLQY), and defect tolerance, making them highly promising for next-generation photonic technologies. In particular, compositional engineering of the halide ions enables systematic tuning of optical properties; however, such approaches often suffer from intrinsic limitations, including phase instability and sensitivity to environmental factors such as moisture, temperature, and illumination. \cite{wang2015transition, kulbak2016cesium, zheng2016improved}

Doping provides an alternative and effective route to tailor the electronic and optical properties of halide perovskites without significantly altering their crystal structure. Incorporation of a small concentration of foreign atoms can substantially modify properties such as photoluminescence, defect densities, and stability. Various dopants, including Mn$^{2+}$, Bi$^{3+}$, and rare-earth ions, have been extensively studied and are known to introduce new emission pathways and enhance optical performance. \cite{parobek2016exciton, liu2016mn2+, hu2017enhancing, bai2018interstitial, yao2018ce3+} In particular, transition-metal doping can also introduce magnetic functionalities, opening avenues for spintronic applications. \cite{amo2010exciton, pradhan2017luminescence, fainblat2017single} Despite these advances, the microscopic mechanisms by which dopants modify the optoelectronic properties of halide perovskites remain incompletely understood.

Among transition-metal dopants, Fe is of particular interest due to its abundance, low cost, and ability to induce rich magneto-optical and catalytic functionalities. For instance, Fe-doped CdS nanocrystals (NCs) exhibit a magneto-optical Stark effect arising from spin-dependent interactions between Fe$^{2+}$ dopant states and host electronic states, leading to magnetically inequivalent excitonic states and concomitant photoluminescence (PL) quenching. \cite{makkar2021magneto} In halide perovskites, Fe has similarly been reported to act as an efficient luminescence quencher, although significantly higher dopant concentrations are required compared to conventional II-VI quantum dots. \cite{chakraborty2022photoluminescence} In addition, Fe doping in CsPbBr$_3$ nanocrystals has been shown to enhance photocatalytic CO$_2$ reduction, promoting selective CH$_4$ evolution over CO, in contrast to the undoped system. \cite{shyamal2019doping} However, these studies do not provide a microscopic understanding of Fe-induced photoluminescence quenching in halide perovskites. In particular, it remains unclear why Fe doping leads to markedly different PL responses in chemically similar systems such as CsPbCl$_3$ and CsPbBr$_3$.

In this work, following earlier experimental observations,\cite{chakraborty2022photoluminescence} we first address the origin of PL quenching in Fe-doped CsPbCl$_3$ using first-principles calculations. We then investigate the role of the halide by extending our study to Fe-doped CsPbBr$_3$ and comparing the PL responses of the two systems. Our experimental results reveal distinct PL behaviour at higher Fe concentrations in the chloride and bromide systems, which cannot be explained solely on the basis of their electronic structures. This motivates us to go beyond an electronic description and examine lattice effects through phonon calculations and electron-phonon coupling analysis. In particular, we analyze electron-phonon coupling to understand the origin of the contrasting PL responses in these materials. Through this study, we aim to develop a unified microscopic understanding of dopant-induced nonradiative processes in halide perovskites. Addressing these questions is essential for designing dopant-engineered materials with controlled nonradiative losses and enhanced optoelectronic performance. Beyond minimizing losses in light-emitting applications, such insights also enable the strategic use of dopants to promote charge separation in photocatalysis and to introduce spin functionality in spintronic devices. Overall, this work provides guiding principles for tailoring nonradiative pathways in halide perovskites, opening avenues for the development of next-generation multifunctional optoelectronic materials.

\section{Computational details} 

All calculations were carried out within the framework of spin-polarized density functional theory (DFT) using the Quantum ESPRESSO package.\cite{giannozzi2009quantum} The Kohn-Sham equations were expanded in a plane-wave basis set with kinetic energy cutoffs of 40 Ry for the wavefunctions and 400 Ry for the charge density. Exchange-correlation effects were treated within the generalized gradient approximation using the Perdew-Burke-Ernzerhof (PBE) functional.\cite{perdew1996generalized} The interaction between valence electrons and ionic cores was described using ultrasoft pseudopotentials.\cite{vanderbilt1990soft}

The pristine CsPbBr$_3$ and CsPbCl$_3$ bulk systems were modeled using five-atom cubic unit cells in the perovskite structure, where each Pb atom is octahedrally coordinated by six halide atoms. For the Fe-doped systems (i.e., Fe-doped CsPbCl$_3$ and Fe-doped CsPbBr$_3$), $2\times2\times2$ supercells were constructed, in which one Pb atom was substituted by Fe to obtain 12.5\% doping, and two Pb atoms were replaced to achieve 25\% doping. The lattice parameters and atomic positions were fully relaxed using the Broyden-Fletcher-Goldfarb-Shanno (BFGS) algorithm\cite{broyden1973local, fletcher1970new, goldfarb1970family, shanno1970conditioning} until all components of all the forces on all atoms were below 0.001 Ry/Bohr. Brillouin zone (BZ) integrations were performed using $8\times8\times8$ Monkhorst-Pack k-point grids\cite{monkhorst1976special} for the primitive cells and $4\times4\times4$ grids for the supercells. Marzari-Vanderbilt cold smearing\cite{marzari1999thermal} with a width of 0.005 Ry was employed to improve convergence. Spin-orbit coupling effects were included using fully relativistic pseudopotentials, given the presence of heavy elements such as Pb and Br.

The vibrational properties were calculated for the optimized 12.5\% Fe-doped CsPbCl$_3$ and 12.5\% Fe-doped CsPbBr$_3$ structures using density functional perturbation theory (DFPT), as implemented in the PHONON module of Quantum ESPRESSO.\cite{giannozzi2009quantum} Dynamical matrices were computed on a $2\times2\times2$ q-point mesh using scalar relativistic ultrasoft pseudopotentials. A stringent self-consistency threshold of $10^{-14}$ was used for the DFPT calculations.


\section{Experimental motivation}

The present combined theoretical and experimental study is motivated by the PL measurements of CsPbCl$_3$ and Fe-doped CsPbCl$_3$ NCs reported by Chakraborty et al.\cite{chakraborty2022photoluminescence}. They systemetically investigated the structural and optical properties using X-ray absorption fine structure (XAFS) spectroscopy together with steady-state absorption and emission measurements. The experiments show that the PL intensity decreases upon Fe doping relative to pristine CsPbCl$_3$, and is further strongly suppressed with increasing Fe concentration, becoming completely quenched at Fe concentrations $\ge 3$\%. These observations naturally lead to the key question addressed in this work: what is the microscopic origin of PL quenching induced by Fe doping in CsPbCl$_3$ NCs?

\begin{figure}[h!]
\centering
    \includegraphics[width=7 cm]{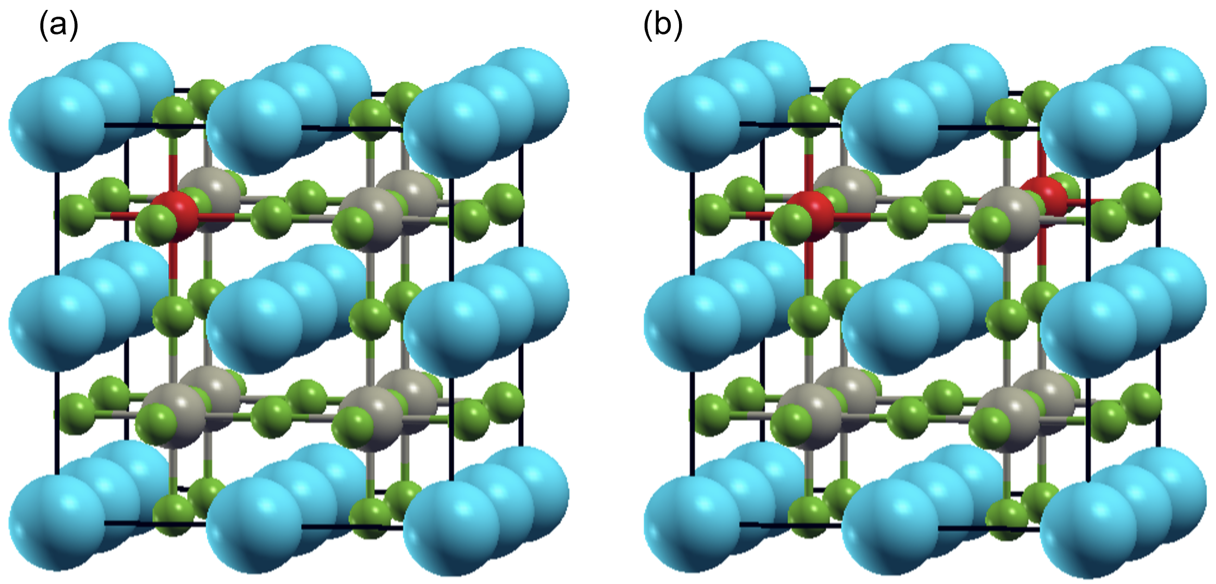}
    \caption{Systems considered in the present DFT calculations: (a) 12.5\% Fe-doped CsPbCl$_3$ and (b) 25\% Fe-doped CsPbCl$_3$. The structures are constructed using a $2\times 2\times 2$ supercell comprising 8 Cs, 7 (6) Pb, 1 (2) Fe, and 24 Cl atoms for 12.5\% (25\%) Fe concentrations. For the 25\% Fe-doped case, only the lowest-energy configuration is shown, in which the two Fe atoms occupy sites along a face diagonal. Color code: Cs (cyan), Pb (gray), Cl (green), and Fe (red).}
    \label{Fig:crystal-structures_Fe-CsPbBr3_and_Fe-CsPbCl3}
\end{figure}

\begin{table}[h!]
\begin{center}
\begin{tabular}{|c|c|c|}
\hline
Fe-Fe arrangement & $\Delta E_{FM}$ (meV) & $\Delta E_{AFM}$ (meV) \\
\hline
1st NN & 397 & 384  \\ \hline
2nd NN & 0 & 36 \\ \hline
3rd NN & 329 & 330  \\ \hline
\end{tabular} 
\end{center}
\caption{DFT-calculated energetics of different configurations of Fe-doped CsPbCl$_3$ at 25\% Fe concentration. The configurations differ in the relative separation between the two Fe atoms as well as in their magnetic ordering (ferromagnetic (FM) or antiferromagnetic (AFM)). $\Delta E$ denotes the energy of each configuration relative to the lowest-energy configuration.}
\label{Tab:Formation-energy_Fe-doped_CsPbCl3}
\end{table}

\section{RESULTS AND DISCUSSION}

\subsection{Optimized crystal structures of CsPbCl$_3$ and Fe-doped CsPbCl$_3$}

We consider bulk cubic CsPbCl$_3$ in the perovskite structure, with a five-atom primitive unit cell consisting of one Cs atom, one Pb atom, and three Cl atoms. To model Fe doping, $2\times2\times2$ supercells are constructed, in which one Pb atom is substituted by Fe atom to obtain 12.5\% Fe doping, and two Pb atoms are replaced to achieve 25\% Fe doping. The corresponding crystal structures for the doped systems are shown in Fig.~\ref{Fig:crystal-structures_Fe-CsPbBr3_and_Fe-CsPbCl3}. The optimized lattice constant of pristine CsPbCl$_3$ is found to be 5.74~\AA, in excellent agreement with previous theoretical results (5.73~\AA)\cite{grote2014tuning}, but slightly larger than experimental values for NCs (5.62~\AA)\cite{chakraborty2022photoluminescence} and bulk (5.61~\AA)\cite{rohrer2001structure}, consistent with the known tendency of the GGA to overestimate lattice parameters.

For the 25\% Fe-doped system, three symmetry-inequivalent configurations are considered, corresponding to different separations between the two Fe dopants: nearest-neighbor (along the lattice vector of distance $a$), next-nearest-neighbor (along the face diagonal, $\sqrt{2}a$), and body-diagonal separation ($\sqrt{3}a$). For each configuration, both ferromagnetic (FM) and antiferromagnetic (AFM) orderings are examined. The relative energetics of these configurations are summarized in Tab.~\ref{Tab:Formation-energy_Fe-doped_CsPbCl3}. We find that the lowest-energy configuration corresponds to Fe atoms occupying next-nearest-neighbor positions along the face diagonal with ferromagnetic ordering. All subsequent calculations for the 25\% Fe-doped system are therefore performed for this ground-state configuration.

The optimized lattice constants decrease systematically with increasing Fe concentration, taking values of 5.629~\AA\ and 5.538~\AA\ for 12.5\% and 25\% Fe doping, respectively. This reduction reflects the smaller ionic size of Fe compared to Pb. A linear dependence of the lattice constant on Fe concentration is observed (see Fig.~2 of the supplementary information (SI)). Furthermore, the calculated formation energies confirm the structural stability of both pristine and Fe-doped systems, with increasing binding strength at higher Fe concentrations (see Tab.~3 of the SI). The monotonic decrease in formation energy with increasing Fe concentration suggests a dopant-induced structural stabilization, where the incorporation of Fe ions shifts the system toward a more energetically favorable equilibrium. This thermodynamic ``deepening" is attributed to the synergistic effect of lattice strain relief and the strengthening of the chemical framework; specifically, the smaller ionic radius of Fe acts as a structural relief valve that allows the halide octahedra to relax into a more ideal geometry, while the 3$d$ orbitals of Fe facilitate stronger, more covalent Fe-Cl bonds compared to the host Pb-Cl bonds in the pristine system.

\subsection{PL quenching in Fe-doped CsPbCl$_3$: Role of midgap Fe $3d$ states}

\begin{figure*}[ht]
    \centering
    \begin{minipage}{0.49\textwidth}
        \centering
        \includegraphics[width=8.5 cm]{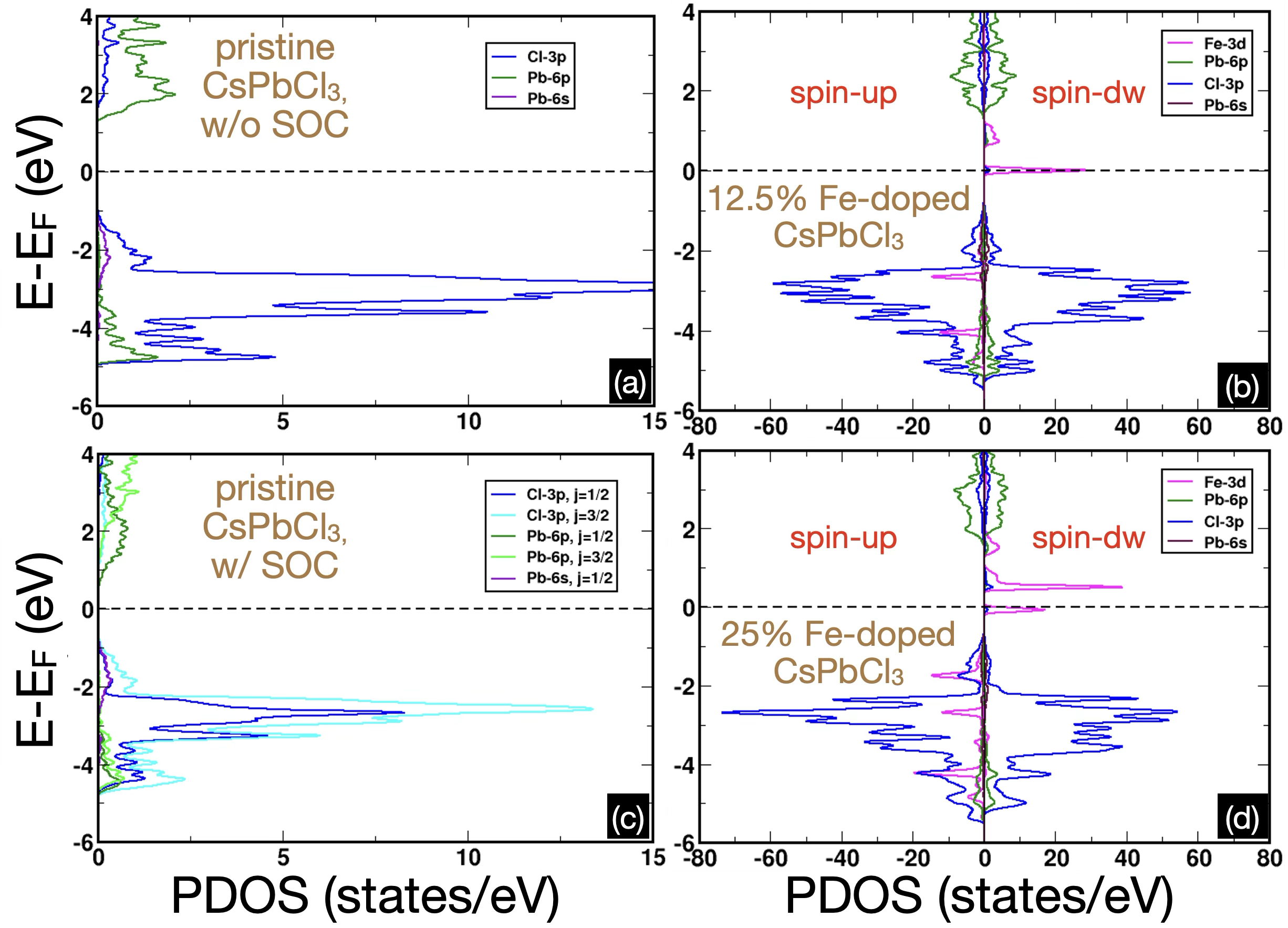}
        \caption{PDOS of pristine CsPbCl$_3$: (a) without SOC and (c) with SOC. Spin-resolved PDOS without SOC for (b) 12.5\% and (d) 25\% Fe-doped CsPbCl$_3$. The electronic states near the Fermi level are dominated by Fe $3d$, Cl $3p$, and Pb $6p$ orbitals.}
        \label{Fig:PDOS_Fe-CsPbCl3_12.5_and_25_woSOC}
    \end{minipage}
    \hfill
    \begin{minipage}{0.49\textwidth}
        \centering
        \includegraphics[width=8.5 cm]{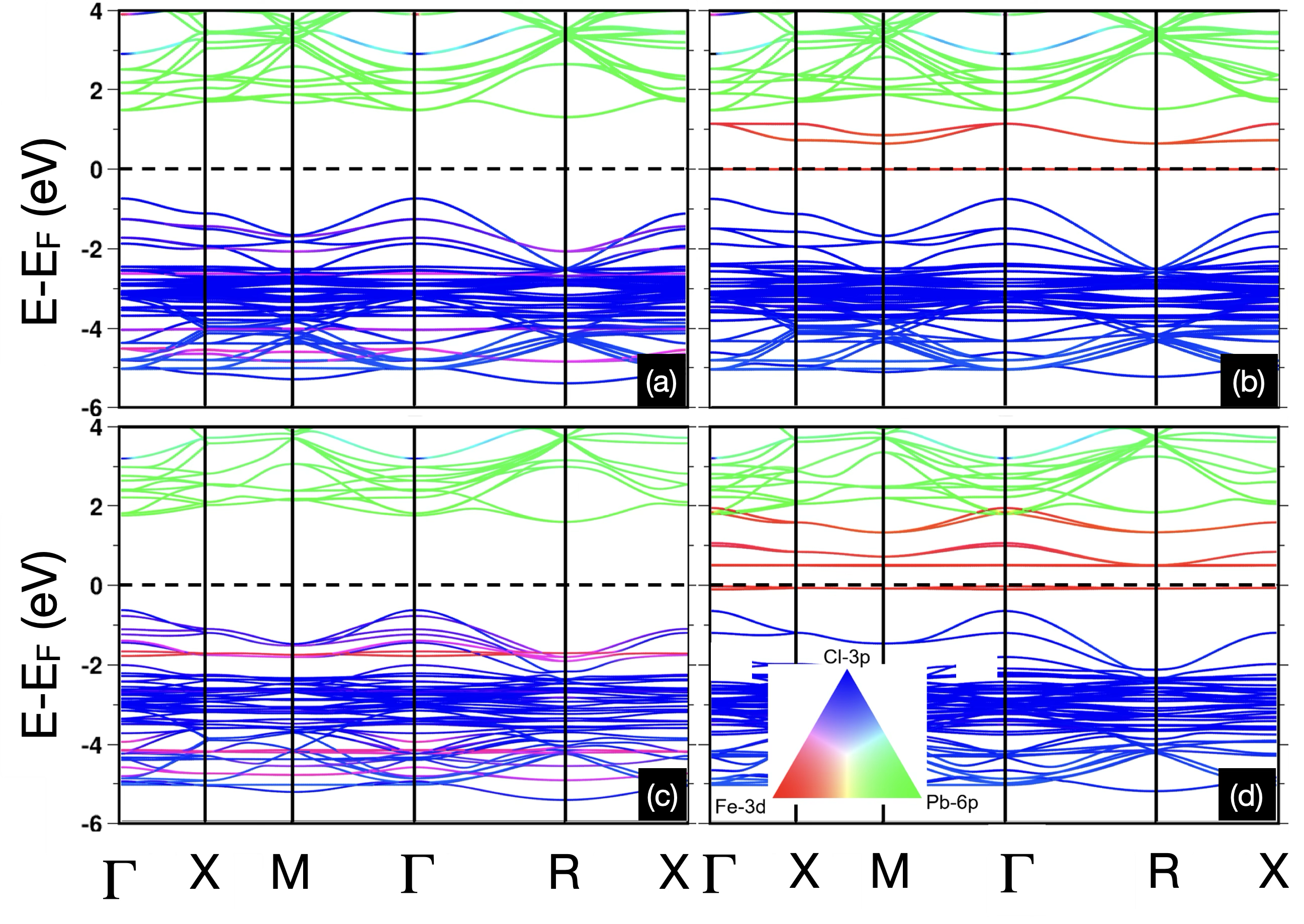}
        \caption{Orbital-projected band structures without SOC for (a) 12.5\% Fe-doped CsPbCl$_3$ (spin-up), (b) 12.5\% Fe-doped CsPbCl$_3$ (spin-down), (c) 25\% Fe-doped CsPbCl$_3$ (spin-up), and (d) 25\% Fe-doped CsPbCl$_3$ (spin-down). The bands are projected onto the three dominant orbitals near the Fermi level: Fe $3d$ (red), Cl $3p$ (blue), and Pb $6p$ (green). The color triangle shown in the inset of panel (d) illustrates the relative orbital contributions, with intermediate colors indicating the degree of orbital mixing (hybridization).}
        \label{Fig:3-color_projected-bands_Fe-CsPbCl3}
    \end{minipage}
    
\end{figure*}

As described under Sect.~III, to understand the PL quenching upon Fe doping in CsPbCl$_3$ NCs, we first examine the orbital-projected density of states (PDOS) of pristine CsPbCl$_3$. Fig.~\ref{Fig:PDOS_Fe-CsPbCl3_12.5_and_25_woSOC}(a) and (c) show the PDOS without and with spin-orbit coupling (SOC), respectively. In both cases, a finite band gap is observed, confirming the semiconducting nature of the material. The calculated PBE-GGA band gap without SOC is 2.22~eV, in excellent agreement with a previous DFT value of 2.21~eV, but smaller than the experimental value of 3.0~eV for bulk CsPbCl$_3$,\cite{gesi1975effect} consistent with the well-known band-gap underestimation in GGA. The PDOS indicates that the valence bands are primarily derived from Cl $3p$ states, while the conduction bands are dominated by Pb $6p$ states. Orbital-projected band structure calculations (see Fig.~4 of the SI) further show that CsPbCl$_3$ is a direct-gap semiconductor, with both the valence band maximum (VBM) and conduction band minimum (CBM) located at the $R$ point of the Brillouin zone. Upon inclusion of SOC [Fig.~\ref{Fig:PDOS_Fe-CsPbCl3_12.5_and_25_woSOC}(c)], splitting of degenerate bands is observed, with a pronounced downward shift of the CBM due to the heavy Pb atom, reducing the band gap to 0.985~eV. This splitting arises from the separation of Pb $6p$ states into $j_{1/2}$ and $j_{3/2}$ components, as seen in the $j$-resolved PDOS.


We next investigate the effect of Fe doping on the electronic structure by substituting Pb atoms with Fe. The PDOS for 12.5\% and 25\% Fe-doped CsPbCl$_3$ without SOC are shown in Fig.~\ref{Fig:PDOS_Fe-CsPbCl3_12.5_and_25_woSOC}(b) and (d), respectively. While the overall band character remains similar to the pristine system, with Cl $3p$ and Pb $6p$ states dominating the valence and conduction bands, respectively, a key modification is the emergence of Fe $3d$ states within the band gap of the pristine system (see the appearance of the magenta-colored peaks within the band gap region of the pristine system). Near the Fermi level, the dominant contributions arise from Fe $3d$ (magenta), Cl $3p$ (blue), and Pb $6p$ (green) orbitals. Unlike pristine CsPbCl$_3$, which is nonmagnetic, Fe doping induces magnetism in the system. Notably, the spin-up channel remains insulating, whereas a sharp and intense peak of Fe $3d$ spin-down states appears in the midgap region and crosses the Fermi level, indicating half-metallic behavior. In addition to this prominent midgap feature, a smaller Fe $3d$ contribution is observed near the CBM. At 25\% Fe concentration, an additional Fe $3d$ peak emerges within the midgap region, accompanied by an overall increase in the spectral weight of Fe $3d$ states due to the higher dopant concentration.

To further elucidate the nature of these states, we analyze the orbital-projected band structure shown in Fig.~\ref{Fig:3-color_projected-bands_Fe-CsPbCl3}, where the bands are resolved into Fe $3d$ (red), Cl $3p$ (blue), and Pb $6p$ (green) contributions. Two distinct types of Fe $3d$ states are identified: (i) highly localized, nearly dispersionless flat bands in the midgap region, corresponding to Dirac delta-like sharp peaks in the PDOS, and (ii) less dispersive Fe $3d$ states located near the CBM, which hybridize with Pb $6p$ states. At higher Fe concentration (25\%), the density of midgap Fe $3d$ states increases, and enhanced hybridization between Fe $3d$ and Pb $6p$ states is observed (see Fig.~\ref{Fig:PDOS_Fe-CsPbCl3_12.5_and_25_woSOC}(d) and \ref{Fig:3-color_projected-bands_Fe-CsPbCl3}(d)).

The implications for PL can be understood in terms of carrier recombination processes. In pristine CsPbCl$_3$, PL arises from radiative recombination of electrons at the CBM with holes at the VBM (see Fig.~4(a) and (c) of the SI). Upon Fe doping, however, the emergence of midgap Fe $3d$ states provides efficient trapping centers for excited electrons (as shown by the red or magenta colored Fe 3$d$ spin-down states in Fig.~\ref{Fig:PDOS_Fe-CsPbCl3_12.5_and_25_woSOC} and \ref{Fig:3-color_projected-bands_Fe-CsPbCl3}). Instead of recombining radiatively across the band gap, electrons relax nonradiatively from the Pb 6$p$ CBM into these midgap Fe 3$d$ states, where their excitation energy is dissipated into lattice vibrations (phonons). Despite non-radiative dominance, residual radiative recombination between excited electrons and holes at the VBM accounts for the low-intensity PL emission. This leads to a significant increase in nonradiative recombination channels and, consequently, a reduction in PL intensity. In this sense, the midgap Fe $3d$ states act as highly efficient carrier traps, effectively behaving as \textit{electronic black holes} for excited carriers. As the Fe concentration increases, the density of midgap Fe $3d$ states also increases (see Fig.~2 and 3), resulting in more efficient trapping of excited carriers and a further enhancement of nonradiative decay processes. This trend explains the experimentally observed progressive quenching of PL with increasing Fe concentration.\cite{chakraborty2022photoluminescence}

An additional important feature is the highly localized nature of the midgap Fe $3d$ states, reflected in their sharp, delta-like peaks in the PDOS and the presence of flat bands in the band structure (see Fig.~\ref{Fig:PDOS_Fe-CsPbCl3_12.5_and_25_woSOC} and \ref{Fig:3-color_projected-bands_Fe-CsPbCl3}). These flat bands correspond to very large effective masses, implying that electrons trapped in these states are highly localized and have negligible mobility. As a result, once carriers are trapped, they are unlikely to return to the conduction band or recombine radiatively with holes in the valence band. Moreover, since these Fe 3$d$ states lie deep within the band gap, they act as efficient \textit{deep trap} centers, further enhancing nonradiative recombination and PL quenching. Thus, the electronic structure provides a clear microscopic explanation for the PL quenching observed upon Fe doping in CsPbCl$_3$ NCs. Further, our results demonstrate that the inclusion of SOC does not qualitatively alter the nonradiative decay mechanisms in Fe-doped CsPbCl$_3$, as discussed above (see Fig.~6 and 8 of the SI).

\subsection{Contrasting PL behavior in Fe-doped CsPbCl$_3$ and Fe-doped CsPbBr$_3$}

\begin{figure}[ht]
\centering
    \includegraphics[width=7.5 cm]{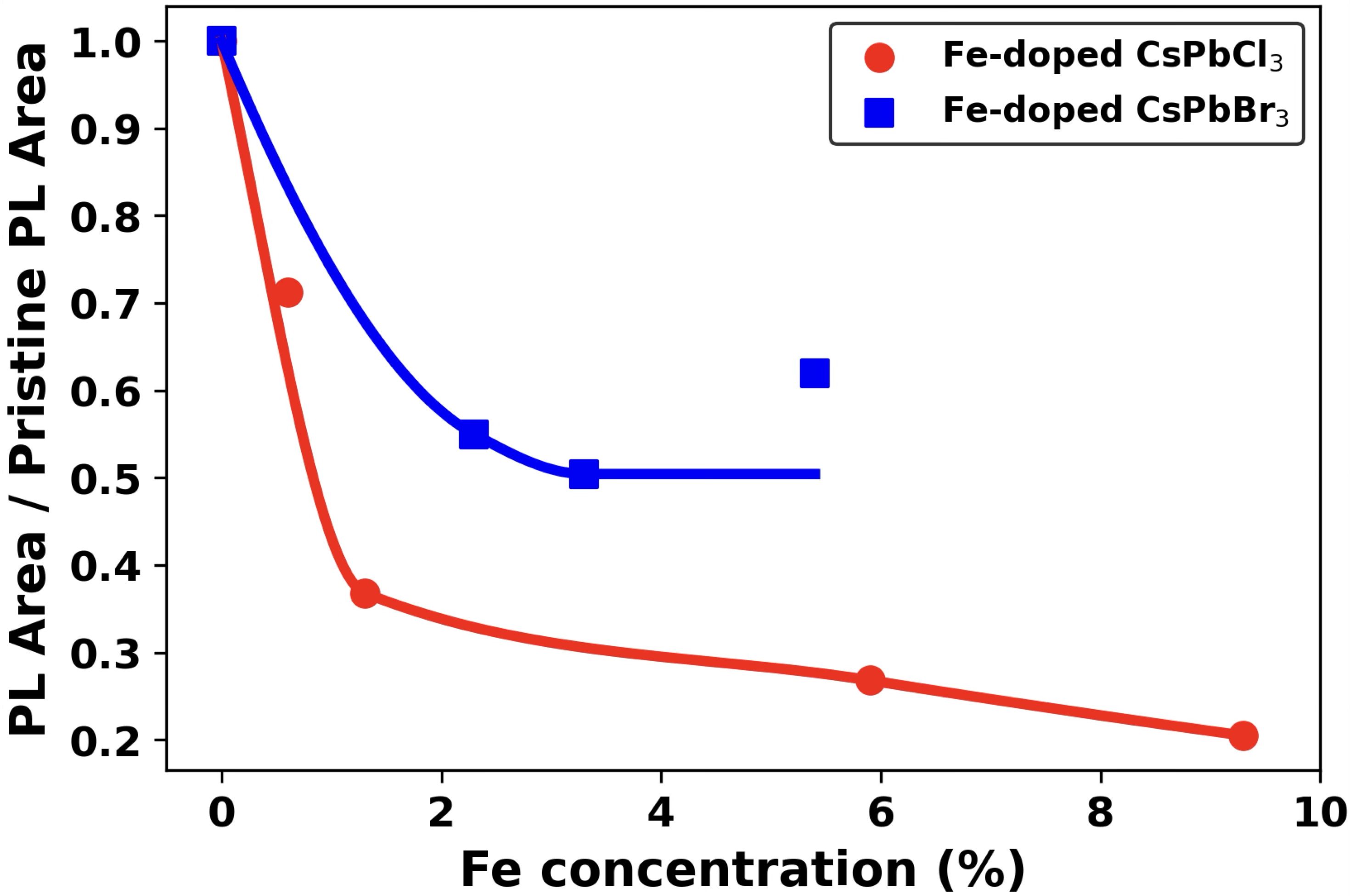}
    \caption{Experimentally measured variation of the integrated PL intensity (area under the emission curve) as a function of Fe concentration for Fe-doped CsPbCl$_3$ (red filled circles) and Fe-doped CsPbBr$_3$ (blue filled squares). The values are normalized with respect to the corresponding pristine systems (0\% Fe doping). The blue and red curves serve as guides to the eye.}
    \label{Fig:PL_Area_comparison}
\end{figure}

Having established that Fe dopants act as electron trap centers, enhancing nonradiative recombination and thereby driving PL quenching, we now turn to the role of the halide. To this end, we investigate the bromide system by measuring the PL intensity of pristine CsPbBr$_3$ and Fe-doped CsPbBr$_3$ at various Fe concentrations. Fig.~\ref{Fig:PL_Area_comparison} shows the integrated PL intensity (area under the PL curve) as a function of Fe concentration for both Fe-doped CsPbCl$_3$ and Fe-doped CsPbBr$_3$. In each case, the PL area is normalized with respect to the corresponding pristine system (0\% Fe doping). Experimental absorbance and PL intensity measurements for varying Fe concentrations in both the chloride and bromide systems are provided in Fig.~9 of the SI. Two key observations emerge from this analysis. First, the introduction of Fe leads to a reduction in PL intensity in both CsPbCl$_3$ and CsPbBr$_3$ relative to their pristine counterparts. Second, the evolution with increasing Fe concentration differs markedly between the two systems: in Fe-doped CsPbCl$_3$, the PL intensity decreases sharply and becomes completely quenched beyond 3\% Fe concentration, whereas in Fe-doped CsPbBr$_3$, the PL intensity decreases initially but then saturates at a finite value for Fe concentrations beyond $\sim$2\%.

To understand the contrasting PL behavior in the chloride and bromide systems, we first examine the electronic structure of Fe-doped CsPbBr$_3$. Our calculations show that both the PDOS and orbital-projected band structure of Fe-doped CsPbBr$_3$ closely resemble those of Fe-doped CsPbCl$_3$ (see Fig.~5-8 of the SI). In particular, Fe $3d$ states emerge in the midgap region, which, as in the chloride system, act as efficient trapping centers for electrons excited to the CBM, thereby enhancing nonradiative recombination and leading to PL quenching upon Fe doping.

Given the close similarity in the electronic structures of Fe-doped CsPbCl$_3$ and Fe-doped CsPbBr$_3$, one would expect a similar evolution of PL intensity with increasing Fe concentration in both systems, ultimately leading to complete quenching. However, this expectation is not borne out experimentally (see Fig.~\ref{Fig:PL_Area_comparison}). While the PL intensity in Fe-doped CsPbCl$_3$ is completely quenched at $\ge 3$\% Fe concentrations, in Fe-doped CsPbBr$_3$ it instead saturates at a finite value at $\ge 2$\% Fe concentration. Thus, despite nearly identical electronic structures, the two systems exhibit markedly different PL responses. This indicates that the change in halide does not significantly modify the electronic structure but has a pronounced effect on the PL behavior. These observations naturally lead to the central question addressed in this work: why does the degree of PL quenching differ between Fe-doped CsPbCl$_3$ and Fe-doped CsPbBr$_3$ at higher Fe concentrations?

\begin{figure}[ht]
\centering
    \includegraphics[width=8.5 cm]{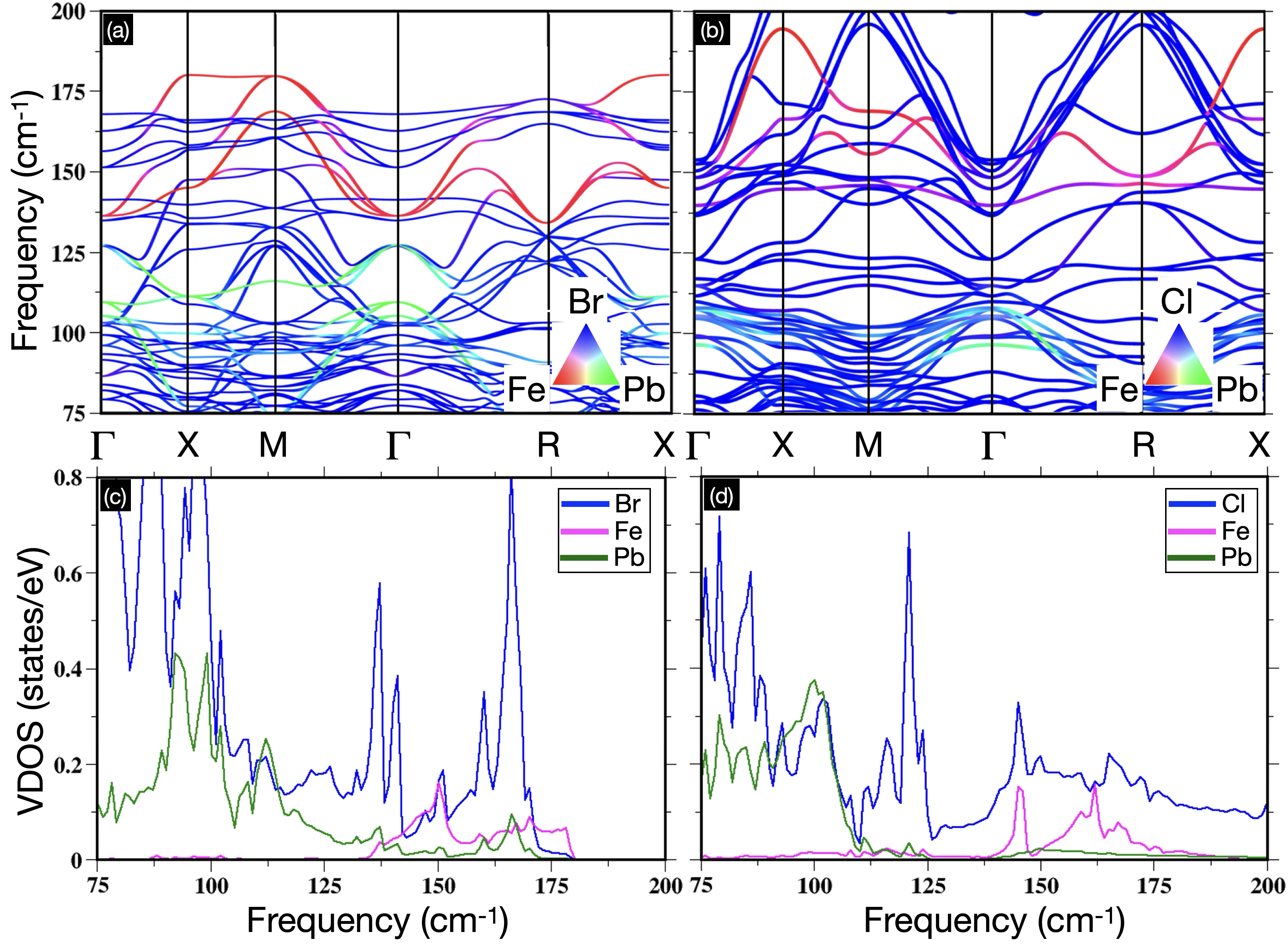}
    \caption{Phonon dispersion (shown in the frequency range of Fe vibrations) with atomic contributions, calculated without SOC for (a) 12.5\% Fe-doped CsPbBr$_3$ and (b) 12.5\% Fe-doped CsPbCl$_3$. The phonon branches are colored according to the relative weights of atomic contributions. The color triangle shown in the inset indicates Fe (red), Br/Cl (blue), and Pb (green), with intermediate colors representing the degree of atomic mixing (hybridization). Phonon density of states (PDOS), also referred to as vibrational density of states (VDOS) (shown in the frequency range of Fe vibrations), with atomic contributions, calculated without SOC for (c) 12.5\% Fe-doped CsPbBr$_3$ and (d) 12.5\% Fe-doped CsPbCl$_3$. Color code: Fe (magenta), Br/Cl (blue), and Pb (green).}
    \label{Fig:phonon-dispersion_and_PDOS_Fe-CsPbBr3_and_Fe-CsPbCl3}
\end{figure}

\subsection{Do phonons alone account for the degree of PL quenching?}

We do not find any significant differences in the electronic structures of Fe-doped CsPbCl$_3$ and Fe-doped CsPbBr$_3$ that can account for the discrepancy observed in their PL responses. The complete quenching of PL at higher Fe concentrations in Fe-doped CsPbCl$_3$ suggests that nonradiative recombination channels dominate entirely, effectively suppressing radiative transitions. This indicates that the excitation energy of electrons is not emitted as photons but is instead transferred to other channels, most likely lattice vibrations. Motivated by this, we next examine the vibrational properties, followed by a detailed analysis of electron-phonon coupling in both Fe-doped CsPbCl$_3$ and Fe-doped CsPbBr$_3$.

The phonon dispersion and atom-projected phonon density of states (PDOS), also referred to as the vibrational density of states (VDOS), including atomic contributions, are shown in Fig.~\ref{Fig:phonon-dispersion_and_PDOS_Fe-CsPbBr3_and_Fe-CsPbCl3}. The vibrational properties are computed using DFPT for the 12.5\% Fe-doped systems only. Due to computational constraints, SOC is not included in these phonon calculations.

To elucidate the role of Fe doping, we focus on the frequency range associated with Fe vibrations, which lies approximately between 135 and 180 cm$^{-1}$ for both the chloride and bromide systems. The phonon dispersion with atomic projections and the corresponding VDOS in this frequency window reveal significant vibrational coupling between Fe and the surrounding halide atoms -- Fe-Br in bromide system and Fe-Cl in chloride system. This is evident from the mixed character of the phonon branches and the overlap of peaks in the PDOS. Such coupling arises naturally from the octahedral coordination of Fe by six halide atoms in the lattice.

In contrast, the contribution of Pb atoms in this frequency range is negligible, as their larger mass leads to vibrational modes predominantly below $\sim$110 cm$^{-1}$. Moreover, due to its lower atomic mass, Cl contributes to higher-frequency vibrations extending beyond 200 cm$^{-1}$, whereas the phonon spectrum of Br terminates at comparatively lower frequencies. This distinction is important and will be revisited below.

Although the Fe-halide vibrational coupling occurs in a similar frequency range for both systems, these results provide an initial indication that the nature and strength of vibrational interactions may differ between the chloride and bromide systems. In particular, the higher vibrational frequencies associated with Cl suggest a greater capacity to absorb electronic excitation energy. This points toward a more efficient conversion of electronic excitation energy into lattice vibrations in the chloride system compared to the bromide counterpart. To quantitatively assess this possibility, we next compute the strength of electron-phonon coupling in Fe-doped CsPbCl$_3$ and Fe-doped CsPbBr$_3$.

\subsection{Role of electron-phonon coupling}

Next, to gain deeper insight into the contrasting PL behavior of Fe-doped CsPbCl$_3$ and Fe-doped CsPbBr$_3$ at higher Fe concentrations, we evaluate the strength of electron-phonon coupling (EPC). To estimate the EPC strength, we follow the approach of X.~Gong \textit{et al.}, and compute the deformation potential, which serves as a quantitative measure of electron-phonon interaction.\cite{gong2018electron} The deformation potential ($D$) is defined as the change in electronic band energy induced by small atomic displacements (e.g., under applied strain), and is given by

\begin{equation}
    D = \frac{\Delta \mathscr{E}}{\Delta l / l_0} \, ,
    \label{Eq:Deformation-potential}
\end{equation}

\noindent where $\Delta \mathscr{E}$ denotes the change in electronic band energy resulting from a small strain $\Delta l$ applied to the optimized lattice parameter $l_0$ of the Fe-doped systems. In the present calculations, we apply both compressive and tensile strains of $-1$\% and $+1$\%, respectively.

\begin{figure*}[ht]
\centering
    \includegraphics[width=17 cm]{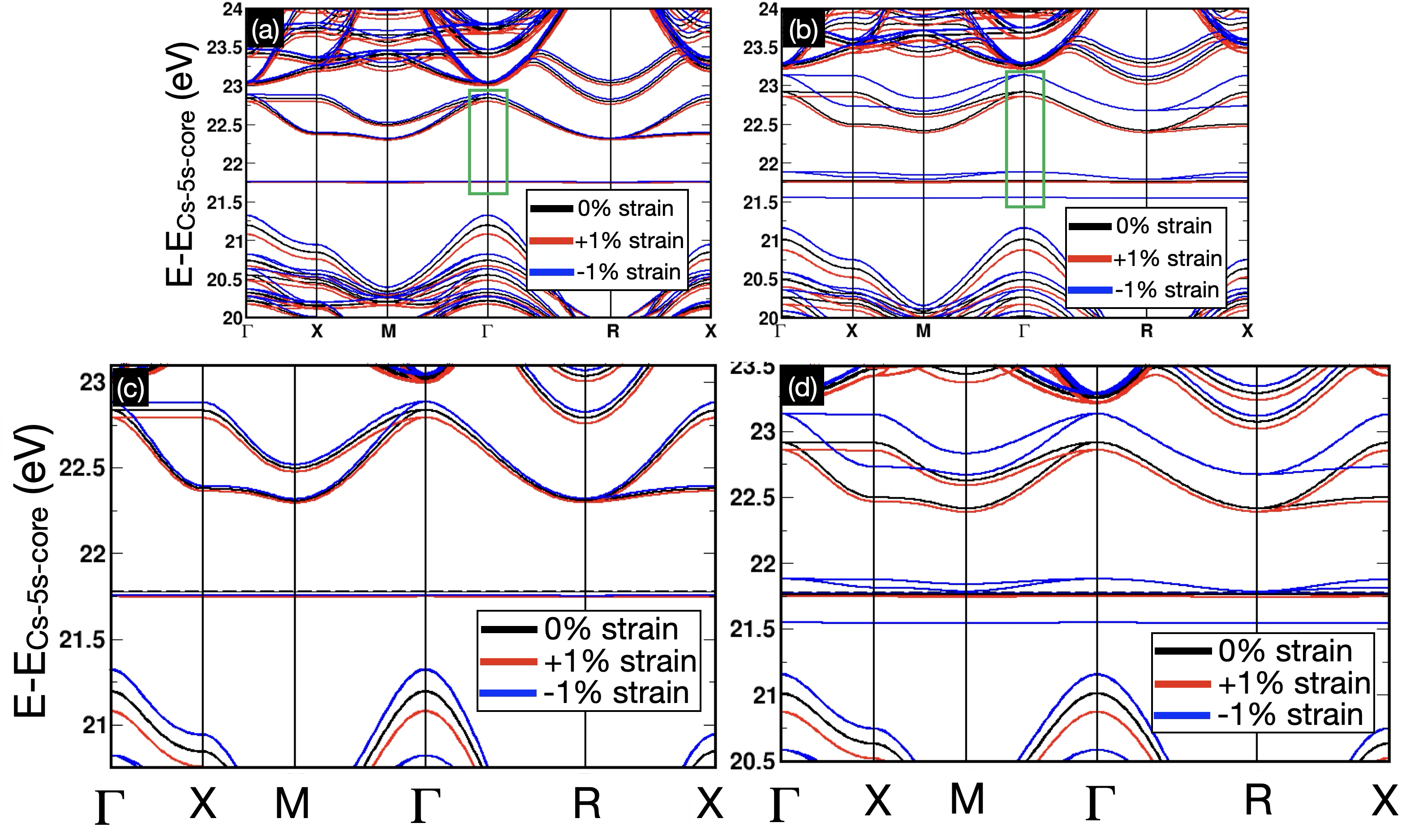}
    \caption{Electronic band structures of (a) 12.5\% Fe-doped CsPbBr$_3$ and (b) 12.5\% Fe-doped CsPbCl$_3$ under applied strain. The corresponding zoomed-in views are shown in (c) and (d), respectively. For direct comparison, band structures for three strain values are overlaid in each panel: 0\% (black), +1\% (red), and $-1$\% (blue).}
    \label{Fig:strain-merged-band-structure_Fe-CsPbBr3_Fe-CsPbCl3}
\end{figure*}

Figures~\ref{Fig:strain-merged-band-structure_Fe-CsPbBr3_Fe-CsPbCl3}(a) and (b) show the electronic band structures of 12.5\% Fe-doped CsPbBr$_3$ and 12.5\% Fe-doped CsPbCl$_3$, respectively, with the band structures under $-1$\% (blue), 0\% (black), and $+1$\% (red) strains superimposed. The corresponding zoomed-in views are presented in Figs.~\ref{Fig:strain-merged-band-structure_Fe-CsPbBr3_Fe-CsPbCl3}(c) and (d). To enable a consistent comparison and isolate strain-induced effects, all band energies are aligned with respect to the Cs $5s$ core level, which is assumed to remain unaffected by strain.

A detailed inspection reveals that the midgap Fe $3d$ states and the VBM exhibit the most significant strain-induced shifts in both systems, whereas the CBM shows comparatively weaker variations. Notably, the strain-induced changes in the VBM and CBM are more pronounced in Fe-doped CsPbCl$_3$ than in Fe-doped CsPbBr$_3$, as will be quantified below. In particular, the midgap Fe $3d$ states are highly sensitive to strain. In Fe-doped CsPbCl$_3$, the triply degenerate flat Fe $3d$ bands split into distinct components under $-1$\% strain, and the weakly dispersive midgap states also exhibit substantial splitting. In contrast, although similar trends are observed in Fe-doped CsPbBr$_3$, the corresponding changes are considerably smaller and less pronounced. For clarity, the region of interest is highlighted by green rectangles in Fig.~\ref{Fig:strain-merged-band-structure_Fe-CsPbBr3_Fe-CsPbCl3}(a) and (b), with the detailed features more clearly visible in the zoomed-in views. Overall, the strain response of the midgap Fe $3d$ states is significantly stronger in the chloride system than in the bromide system.

To quantify the strain-induced changes in the electronic structure, we select representative $k$-points and bands that are most relevant for band-edge transitions and carrier trapping: (i) the VBM at the $\Gamma$ point, (ii) the CBM at the $R$ point, (iii) the extrema (maxima and minima) of the flat Fe $3d$ bands, and (iv) the extrema of the weakly dispersive Fe $3d$ bands. For each of these states, we compute the change in electronic band energy under applied strain and evaluate the deformation potential using Eq.~\ref{Eq:Deformation-potential}. The shift in electronic energy for various strain values in both Fe-doped CsPbCl$_3$ and Fe-doped CsPbBr$_3$ is presented in Fig.~10 of the SI. The final value of $D$ is obtained by averaging over these selected states. Our calculations yield $D = 3.71$ eV for 12.5\% Fe-doped CsPbBr$_3$ and $D = 10.55$ eV for 12.5\% Fe-doped CsPbCl$_3$, indicating that the deformation potential -- and hence the EPC strength -- is larger by a factor of 2.84 in the chloride system compared to the bromide system.

To further estimate the EPC matrix element, the elastic constant $C_{11}$ is required. We approximate the elastic constants of the Fe-doped systems by those of the corresponding pristine systems, which is justified given the relatively low Fe concentrations in experiments. Accordingly, we compute $C_{11}$ for pristine CsPbBr$_3$ and CsPbCl$_3$ using the following relations for cubic systems:\cite{mehl1990structural}

\begin{equation}
    \Delta E_{\mathrm{tot}} = 3V (C_{11} - C_{12}) e_1^2 + O(e_1^3) \, ,
    \label{Eq:elastic-constant-1}
\end{equation}

and

\begin{equation}
    3B = C_{11} + 2C_{12} \, ,
    \label{Eq:elastic-constant-2}
\end{equation}

\noindent where $\Delta E_{\mathrm{tot}}$ is the change in total energy of the strained system relative to the unstrained configuration, $V$ is the optimized unit cell volume, $e_1$ is the applied strain, and $B$ is the bulk modulus. We consider a set of strained configurations obtained by applying both tensile and compressive strains while keeping the volume fixed at $V$. The $\Delta E_{\mathrm{tot}}$ versus $e_1$ data are fitted to a third-order polynomial, which is effectively quadratic due to the negligible contribution of higher-order terms. From the fitting parameters, we extract $C_{11} - C_{12}$ using Eq.~\ref{Eq:elastic-constant-1}. The value of $C_{11} + 2C_{12}$ is obtained from the bulk modulus via Eq.~\ref{Eq:elastic-constant-2}. Solving these equations yields $C_{11} = 42.42$ GPa for CsPbBr$_3$ and $C_{11} = 49.15$ GPa for CsPbCl$_3$. 

The probability of an electron undergoing a transition from an initial state $\mathbf{k}$ (this represents plane waves for a free electron in the CBM) to a final state $\mathbf{k'}$ via interaction with phonons arising from lattice vibrations is governed by squared modulus of the matrix element:\cite{bardeen1950deformation}

\begin{equation}
    |M|^2 = \frac{k_B T D^2}{V C_{11}} \, ,
    \label{Eq:transition-probability_EPC}
\end{equation}

\noindent where $k_B$ is the Boltzmann constant, $T$ is the temperature, $V$ is the crystal volume, $C_{11}$ is the elastic constant, and $D$ is the deformation potential computed above. This expression shows that the transition probability scales with the square of the deformation potential and inversely with the elastic constant.\cite{bardeen1950deformation} This formalism describes electron transitions induced by lattice deformations, i.e., in the presence of phonons generated by atomic displacements from equilibrium. In the present context, the transition of an electron from the CBM (initial state $\mathbf{k}$) to the midgap Fe $3d$ states (final state $\mathbf{k'}$) is mediated by phonons. Therefore, the transition probability provides a direct measure of the EPC strength, implying $\mathrm{EPC} \propto \frac{D^2}{V C_{11}}$. 

We compute the squared modulus of the transition matrix element $|M|^2$ using Eq.~\ref{Eq:transition-probability_EPC}, taking $T = 300$ K (room temperature) and $V$ as the unit cell volume of the respective systems. It should be noted that the deformation potential $D$ and the elastic constant $C_{11}$ have been determined in the preceding analysis. Our results yield $|M|^2 = 2.1 \times 10^{-45}$ for Fe-doped CsPbBr$_3$ and $|M|^2 = 17 \times 10^{-45}$ for Fe-doped CsPbCl$_3$ (both in SI units). This indicates that the EPC strength, as reflected in $|M|^2$, is larger by a factor of $\sim$8.1 in the chloride system compared to the bromide system. This is a key result, demonstrating that the probability of an excited electron transitioning from the CBM to the midgap Fe $3d$ states via phonon-mediated processes is significantly higher in Fe-doped CsPbCl$_3$ than in Fe-doped CsPbBr$_3$. Such transitions are nonradiative, as the electronic excitation energy is transferred to lattice vibrations rather than being emitted as photons. Consequently, upon Fe doping, a fraction of excited electrons in both systems undergo nonradiative recombination via midgap Fe $3d$ states, while the remaining electrons recombine radiatively with holes at the VBM, giving rise to PL. This competition leads to an overall reduction in PL intensity relative to the pristine systems. However, the extent of this reduction differs markedly between the two materials.

In Fe-doped CsPbCl$_3$, the lighter Cl atoms support higher-frequency lattice vibrations (extending beyond $\sim$175 cm$^{-1}$, as shown in Fig.~\ref{Fig:phonon-dispersion_and_PDOS_Fe-CsPbBr3_and_Fe-CsPbCl3}(b) and (d)), resulting in a larger population of phonons and enhanced electron-phonon interactions. As a result, nearly all excited electrons are funneled into nonradiative channels, leading to complete quenching of PL at higher Fe concentrations. In contrast, in Fe-doped CsPbBr$_3$, the heavier Br atoms vibrate at lower frequencies and generate relatively fewer number of phonons. This results in weaker EPC, such that only a fraction of excited electrons undergo nonradiative transitions (from the CBM to the midap Fe 3$d$ states), while the remaining carriers recombine radiatively (with the holes in the VBM). Consequently, a finite PL intensity persists even at higher Fe concentrations in the bromide system.

Therefore, the significantly larger EPC in Fe-doped CsPbCl$_3$ leads to a higher rate of nonradiative recombination compared to Fe-doped CsPbBr$_3$, resulting in complete quenching of PL at higher Fe concentrations in the chloride system. Despite the two materials exhibiting nearly identical electronic structures, including the presence of midgap Fe $3d$ trap states, the markedly different PL responses can be attributed to the disparity in their electron-phonon coupling strengths. We thus argue that the contrasting PL behaviour at higher Fe concentrations originates from the more efficient conversion of electronic excitation energy into lattice vibrational energy in Fe-doped CsPbCl$_3$ than in Fe-doped CsPbBr$_3$, which enhances nonradiative decay channels and ultimately leads to fully quenched PL in the chloride system.


\section{Summary and Conclusions}

In summary, we have carried out a comprehensive theoretical investigation to uncover the microscopic origin of PL quenching in Fe-doped CsPbX$_3$ (X = Cl, Br) nanocrystals. Our electronic structure calculations reveal that Fe incorporation introduces midgap Fe $3d$ states, which act as efficient trap centers for excited electrons and promote nonradiative recombination, thereby explaining the initial reduction of PL intensity upon Fe doping in CsPbCl$_3$. Extending our study to CsPbBr$_3$, we find that despite exhibiting nearly identical electronic structures, including similar midgap Fe $3d$ states, the two systems display markedly different PL behaviour at higher Fe concentrations: complete quenching in the chloride system and saturation at a finite value in the bromide system. This indicates that electronic structure alone is insufficient to account for the contrasting PL responses. Phonon analysis suggests comparable vibrational features in both systems, but does not provide a quantitative explanation. By explicitly evaluating the electron-phonon coupling strength through deformation potential analysis, we demonstrate that EPC is significantly stronger in Fe-doped CsPbCl$_3$ than in Fe-doped CsPbBr$_3$. In particular, the broader spectrum of phonon frequencies associated with the lighter Cl atoms enhances the interaction between excited electrons and phonons, enabling more efficient dissipation of electronic excitation energy into lattice vibrations in the Fe-doped chloride system. As a result, nonradiative recombination channels dominate in the chloride system, leading to fully quenched PL at higher Fe concentrations, whereas weaker EPC in the bromide system allows a fraction of carriers to recombine radiatively, resulting in a finite PL intensity. Our results establish that, beyond electronic structure, the strength of electron-phonon coupling -- governed by the nature of the halide ions -- plays a decisive role in determining the extent of PL quenching in Fe-doped halide perovskites. This work provides fundamental insights into dopant- and lattice-driven nonradiative processes and offers a guiding framework for tailoring the optoelectronic performance of doped perovskite materials.

\section{Acknowledgements} 

A.D. acknowledges JNCASR and the Department of Science and Technology (DST), Government of India, for financial support through his Ph.D. and postdoctoral fellowships. He also acknowledges the Sheikh Saqr Laboratory at ICMS, JNCASR, TUE-CMS, JNCASR, and the Param Yukti supercomputer at JNCASR under the National Supercomputing Mission (NSM) for providing computational resources. S.C. gratefully acknowledges financial support through the DST-INSPIRE doctoral fellowship from the DST, Government of India. A.D. sincerely thanks Dr. Debdipto Acharya for fruitful discussions and valuable assistance with technical calculations.

\bibliography{reference} 

\end{document}